\documentstyle[12pt]{article}   
\begin{document}

\begin{center}
{\large{\bf Dissipative tunneling in presence of classical chaos in a mixed
quantum-classical system}}
\end{center}

\begin{center}
{\bf Bidhan Chandra Bag$^1$, Bikash Chandra Gupta$^2$ and Debshankar Ray$^1$ \\
$^1$Indian Association for the Cultivation of Science \\
Jadavpur, Calcutta 700 032, INDIA. \\
$^2$Institute of Physics, Sachivalaya Marg, \\
Bhubaneswar 751 005, INDIA.} 
\end{center}

\vspace{0.5cm}

\begin{abstract}
We consider the tunneling of a wave packet through a potential barrier which
is coupled to a nonintegrable classical system and study the interplay of
classical chaos and dissipation in the tunneling dynamics. We show 
that chaos-assisted tunneling is further enhanced by dissipation, while
tunneling is suppressed by dissipation when the classical subsystem is regular.

\end{abstract}

\vspace{0.2cm}

PACS number (s) :05.45.+b, 03.65 Bz

\newpage

\begin{center}
{\bf{I. \hspace{0.2cm}Introduction}}
\end{center}
The systems with mixed quantum-classical description have been the subject of
considerable interest in recent years [1-9]. The validity of this description 
essentially rests on whether the quantum effects of one subsystem is 
negligible compared to the others. The well-known example is the Maxwell-Bloch
equations [3, 4] which describe a model two-level system (quantum mechanical) 
interacting with a strong single mode classical electromagnetic field. The
others comprise the models involving nuclear collective motion [5], one 
dimensional molecule [6] where the 
motion of an electron is described by an effective potential provided by the 
nuclei and other electrons within adiabatic approximation scheme, etc. The mixed
description has also been employed by Bonilla and Guinea [7] for the measurement 
processes. Taking recourse to an average description in terms of an effective
classical Hamiltonian, Pattanayak and Schieve [8] have studied some interesting
aspects of semi-quantal chaos.

Very recently the interplay of classical and quantum degrees of freedom in a 
special class of systems with system-operators pertaining to a closed Lie 
algebra with respect to system Hamiltonian has been demonstrated to have 
acquired special relevance in connection
with dissipation in quantum evolution[9]. While the quantum degree of freedom is responsible
for the generic quantum features the implication of classicality is two-fold. 
First, it has been shown
that if the classical degree of freedom is assigned the type of 
evolution as prescribed by the classical treatment of dissipative systems it is 
possible to realize dissipation for such composite systems via an indirect route without any 
violation of quantum rule. 
Second, if the classical subsystem is nonintegrable then the quantum motion by virtue of
this nonintegrability may be profoundly affected. 
Our object here is to study this dissipative
evolution of a quantum system coupled to a nonintegrable classical system.

Before going into the further details let us elaborate this issue a little 
further in somewhat general terms.

We first consider the coupling of a quantum system to a classical system in 
terms of the following Hamiltonian
\begin{equation}
\hat{H} = \hat{H}_{q} + H_{cl} + \hat{H}_{q-cl}   \; \; \;, 
\end{equation}
\noindent
where $\hat{H}_{q}$ and $H_{cl}$ refer to quantum and classical subsystems of 
the total Hamiltonian, respectively. $\hat{H}_{q-cl}$ is the interaction 
potential which contains both classical and quantum degrees of freedom.
If the quantum operators $\left\{\hat{R}_{i}\right\}$ form a closed algebra with
respect to the Hamiltonian $\hat{H}$ of the system, i. e. , if we have 
relations of the type
\begin{equation}
\left[ \hat{H}(t), \hat{R}_{i} \right ] = i\hbar \sum_{j=1}^n g_{ji}(t) \hat{R}_{j} 
\; \; \; i=1,2...n \; \;,
\end{equation}
\noindent
where $g_{ji}$ are the elements of a $n \otimes n$ matrix, then one can have, by virtue 
of the equations of motion
\begin{equation}
\frac{d \hat{R}_{i}}{dt} = \frac{i}{\hbar} \left[ \hat{H}, \hat{R}_{i} \right] \; \; \; \; i=1,2...n \; \;,
\end{equation}
a set of linear first order differential equations
\begin{equation}
\frac{d \langle \hat{R}_{i} \rangle}{dt} = -\sum_{j=1}^n g_{ji} \langle \hat{R}_{j} \rangle \;,
\end{equation}
\noindent
which describes the temporal evolution of the expectation values.

It is important to note that Eq. (4) depends crucially on the elements $g_{ji}$
which are determined by the classical dynamical variables as appeared in the
Hamiltonian $\hat{H}$ (Eq.1). In other words, $g_{ji}$ is dependent on classical
subsystem $H_{cl}$. If the energy is taken to be the quantum expectation 
value of the Hamiltonian $\hat{H}$, then one can easily generate the temporal
evolution of classical conjugate variables, coordinate $x$ and momentum $p$
using $\langle \hat{H} \rangle$ as follows;
\begin{eqnarray}
\frac{dx}{dt} & = & \frac{\partial \langle \hat{H} \rangle}{\partial p} \;, \\
\frac{dp}{dt} & = & -\frac{\partial \langle \hat{H} \rangle}{\partial x}  \;.
\end{eqnarray}
\noindent
Eqs. (4-6) comprise the complete dynamical evolution corresponding to 
Hamiltonian $\hat{H}$ in the dissipation-free case.

We now take into consideration two different aspects of classical motion of the
subsystem $H_{cl}$.

First, the classical dynamical variable is made dissipative in an ad hoc fashion 
by adding $- \gamma p$ to Eq.(5), i. e. ,
\begin{equation}
\frac{dp}{dt} = -[ \frac{\partial \langle \hat{H} \rangle}{\partial x} +\gamma p] \; \; \;.
\end{equation}
\noindent
This allows [9] an indirect route to  dissipation in the quantum evolution
because of $\hat{H}_{q-cl}$ term in Eq.(1). It is easy to check that this 
quantum description is quite valid since  no quantum rule is violated in the 
process. Thus Eqs. 4, 5 and 7 govern the disspative dynamical evolution. We note, in 
passing, that this method has a distinct advantage over some other approaches [10-12] 
to quantum theory of damping which relies on explicit time-dependent Hamiltonian,
such as, Kanai Hamiltonian [10] etc., since these approaches lead to contradiction
with uncertainty principle.

The second aspect of classicality of the subsystem concerns the nature of 
motion generated by $H_{cl}$. If the classical subsystem is non-integrable
then the quantum evolution is affected through the properties of $g_{ji}$ 
matrix. It is important to emphasize a pertinent point at this stage. There are 
well-known cases where exponential instability occurs in a system where the overall
dynamical system is non-integrable. For example, we refer to the case of 
optical chaos described by Maxwell-Bloch equations [3, 4]. Similar consideration of
quantum-classical mixed description has been given to explore quantum chaos 
within Born-Oppenheimer approximation in a typical molecular system where the fast electronic motion and 
the  slow nuclear motion (classical) separate in a very natural way.
In contrast to these cases the present paper concerns the 
nonintegrability due to the subsystem $H_{cl}$ only.

The studies of dissipative effects in quantum systems are usually based on a
traditional system-reservoir linear coupling model. There are two standard
ways of formulating the problem[13-15, 21, 25]. The first one is based on density operator
approach and leads to the socalled master equation while the second one is
based on the Heisenberg picture and leads to the noise operators. In the latter
case one replaces the reservoir by damping terms in the  Heisenberg equations
of motion for a dissipation-free system and adds fluctuating forces as the 
driving terms which impart fluctuations to the system. The operator forces 
are such that, first, the system has the correct statistical properties to 
agree in the classical limit and second, they maintain the commutation 
properties of the Boson operators to ensure that uncertainty principle is
not violated. These consideration are taken care of in the treatments of 
density based or operator based theories (e. g. , see Leggett and co-workers[13]
and Krive and coworkers [14] and others [15]. The spiritual root of quantum
statistical approach to damping lies in the fluctuation-dissipation theorem, 
which illustrates a dynamical balance of inward flow of energy due to
fluctuations from the reservoir into the system and the outward flow of energy
from the system to the reservoir due to dissipation of the system mode.
Such a dynamical balance is automatically maintained 
in the present treatment (and one need not add a stochastic term in Eq.(7)) 
through
a feedback from quantum to classical subsystem since here one deals with
a finite number of degrees of freedom in the quantum plus classical subsystems
as compared to the infinite number of degrees of freedom of the reservoir in 
the traditional system-reservoir description. The closure property of the 
algebra pertaining to the quantum system thus plays an important role in the 
feedback mechanism. Apart from being simple to implement a decisive 
advantage of the approach is that it may be carried over to the nonlinear
systems (e. g. , a Morse oscillator described by SU(2) or SU(1, 1) Lie algebra [16])
in a straight forward manner,
while the master equations are strictly valid for the linear models.

The consideration of the two  aspects of classicality as mentioned earlier 
therefore 
leads us to dissipative chaotic evolution of classical degrees of freedom.
Our object is to explore the influence of classical chaos on 
quantum evolution, dissipation being realized indirectly in the overall dynamics
through
classical friction. The generic quantum feature that we study here is 
tunneling. It is extremely important to asses what influence if any classical
chaos have on it, particularly in presence of dissipation. Although the 
interplay of chaos and tunneling [17-19] or tunneling and dissipation [14, 20, 21] have been the
subject of many researchers separately over the last decade, it is not 
clear how dissipative tunneling is influenced by classical chaos. We are 
therefore concerned here with all three interplaying aspects of evolution, 
e. g. , classical chaos, dissipation and tunneling in a typical model
with mixed quantum-classical description.

\vspace{0.5cm}
 
\begin{center}
{\bf
II. \hspace{0.2cm} The model and the mixed quantum-classical dynamics}
\end{center}
We consider a particle with fixed energy $E_q<0$ penetrating an inverted 
potential barrier. $-E_q$ is the energy measured from the top of the barrier 
(Fig. 1). The quantum subsystem $\hat{H}_{q}$ which describes the penetration of
the particle through the barrier is given by
\begin{equation}
\hat{H}_{q} = \frac{\hat{p}_{q}^2}{2m}-\frac{1}{2}m \omega_{0}^2 \hat{x}_{q}^2 \; \;\;.
\end{equation}

\noindent
$\hat{x}_{q}$ and $\hat{p}_{q}$ are the quantum mechanical operators 
corresponding to position and momentum of the particle respectively. m is the mass of the
particle and $\omega_{0}$ refers to the frequency of the inverted well.

The Hamiltonian for the classical subsystem is given by
\begin{equation}
H_{cl} = \frac{p_{cl}^2}{2M} + a x_{cl}^4 - b x_{cl}^2 + 
g x_{cl} \; \cos \omega_1t \; \; \;,                        
\end{equation}
\noindent
which governs the motion of a classical system with mass $M$ and characterized
by its position $x_{cl}$ and momentum variable $p_{cl}$ in a double-well 
potential driven by an external field with frequency $\omega_{1}$. $a$ ad 
$b$ are the parameters of the double-well oscillator. $g$ includes the effect of
coupling of the external field to the oscillator and the strength of the field.

$H_{cl}$ is nonintegrable and has been widely employed by various workers [17-19, 22-23] over 
the last few years in a variety of situations related to classical and quantum 
chaos.

The description of the Hamiltonian with mixed degrees of freedom is now made 
complete by considering the coupling of classical and quantum degrees of 
freedom in terms of the interaction potential $\hat{H}_{q-cl} (= \lambda x_{cl}
\hat{x}_{q})$ so that we have 
\begin{equation}
\hat{H} = \hat{H}_{q} + H_{cl} + \lambda x_{cl} \hat{x}_{q}   \; \; \;,
\end{equation}
\noindent
$\lambda$ represents the coupling constant.

Making use of the following rescaled dimensionless quantities
\begin{equation}
\left. \begin{array}{ccccccc}
\hat{x} & = & (m \omega_{0})^\frac{1}{2} \hat{x}_{q} &, &
\chi & = & \frac{\lambda}{(mM \omega \omega_{0})^\frac{1}{2}} \\
& & & & & &  \\
\hat{p} & = & (m \omega_{0})^{- \frac{1}{2}} \hat{p}_{q} &, &
A & = & \frac{a}{M^2 \omega^3} \\
& & & & & & \\
s & = & (M \omega)^\frac{1}{2} x_{cl} &, &
B & = & \frac{b}{M \omega^2} \\
& & & & & & \\
p_{s} & = & (M \omega)^{- \frac{1}{2}} p_{cl} &, &
G & = & \frac{g}{(M \omega^3)^\frac{1}{2}}  \\
\end{array} \right \} \; \; \; ,
\end{equation}

\noindent
we rewrite the Hamiltonian(10)
\begin{equation}
\hat{H} = \omega_{0} \frac{\hat{p}^2}{2} - \omega_{0} \frac{ \hat{x}^2}{2}
+ \frac{\omega p_{s}^2}{2} +\omega A s^4 -B \omega s^2  
+ G \omega s \cos{\omega_{1} t}  + \chi \omega_{0} s \hat{x} \; \; \;,
\end{equation}
\noindent
where $\omega$ is the linearized frequency of the double-well potential.

By choosing the relevant operators belonging to the set $\{ \hat{1}, \hat{x},
\hat{p}, \hat{x^2}, \hat{p^2}, \hat{L} = \hat{x} \hat{p} + \hat{p} \hat{x} \}$
we may construct a partial Lie algebra ($\hat{1}$ is the unity operator).
The temporal evolution of the expectation values of these operators (see Eq.4)
is given by the following set of coupled differential equations
\begin{eqnarray}
\frac{d \langle \hat{x} \rangle}{d \tau} & = & \langle \hat{p} \rangle \; , 
\nonumber\\ 
\frac{d \langle \hat{p} \rangle}{d \tau} & = & \langle \hat{x} \rangle 
- \chi s \; , \nonumber\\ 
\frac{d \langle \hat{x^2} \rangle}{d \tau} & = & \langle \hat{L} \rangle \; ,
\nonumber\\ 
\frac{d \langle \hat{p^2} \rangle}{d \tau} & = & \langle \hat{L} \rangle 
- 2 \chi s \langle \hat{p} \rangle \; , \nonumber\\ 
\frac{d \langle \hat{L} \rangle}{d \tau} & = & 2(\langle \hat{p^2} \rangle 
+ \ \langle \hat{x^2} \rangle - \chi s \langle \hat{x} \rangle) \; ,
\end{eqnarray}
\noindent
where $\tau$ denotes the scaled dimensionless time $\tau = \omega_{0} t$.

Since $\langle H \rangle$ coincides with the energy of the system, the 
classical equations of motion are
\begin{eqnarray}
\frac{d s}{d \tau} & = & \Omega p_{s} \; , \nonumber\\ 
\frac{d p_{s}}{d \tau} & = & -\left[ \chi \langle \hat{x} \rangle + 
G \Omega \cos \Omega_1 \tau  
+ 4 A \Omega s^3 -2B \Omega s+ \Gamma p_{s} \right]   \; \; \; ,
\end{eqnarray} 
\noindent
where,
\vspace{-0.5cm}
\begin{eqnarray*}
\Omega & = & \frac{\omega}{\omega_{0}} \; , \nonumber\\ 
\Omega_{1} & = & \frac{\omega_{1}}{\omega_{0}} \; , \nonumber\\ 
\end{eqnarray*}
\vspace{-2.5cm}

\noindent
and
\begin{equation}
\Gamma = \frac{\gamma}{\omega_{0}}  \; \; \;.
\end{equation}
\noindent
Here $\Gamma$ as defined above is the rescaled dimensionless damping constant
introduced in the classical equations of motion (14) in an ad hoc fashion.
We have already pointed out that this ad hoc introduction of classical friction
leads to dissipation in the overall dynamics without any contradiction to uncertainty principle.
Eqs. (13) and (14) thus govern the complete dynamics of the mixed system.

We now turn to the motion of the wave packet. In order that the motion of a 
wave packet comes close to the motion of a classical particle it is necessary 
that its average position and momentum follow the laws of classical mechanics.
However, this condition is automatically satisfied for the inverted harmonic
potential barrier we consider here.

We now describe the wave function of a particle by a Gaussian wave packet of the
form
\begin{equation}
\psi(x,t) = N(t) \exp{\left[- \beta(t) (x- \alpha(t))^2 \right]}  \; \; \; ,
\end{equation}
\noindent
where $\alpha$(t) and $\beta$(t) are two complex, time-dependent parameters 
to  be determined. $N$(t) is a normalization factor which is not of much 
interest here. The Gaussian wave packets have the decisive advantage since the  
ansatz (16) is a solution of the Schrodinger equation,
\begin{equation}
\left(\hat{H} -i \frac{\partial}{\partial t} \right) \psi(x,t) = 0  
\end{equation}
\noindent
if $\alpha$ and $\beta$ satisfy the following two equations
\begin{eqnarray*}
i\frac{d \beta}{dt} = 2 \omega_{0} \beta^2 + \frac{\omega_{0}}{2}
\end{eqnarray*}
\noindent
and
\begin{equation}
i\beta \frac{d \alpha}{dt} = \chi \frac{\omega_{0} s(t)}{2} - \frac{\omega_{0} \alpha}{2}  
\end{equation}

or their scaled version (using $\tau = \omega_{0} t$)
\begin{eqnarray}
i\frac{d \beta}{d \tau} & = & 2  \beta^2 + \frac{1}{2} \; , \nonumber\\ 
i\beta \frac{d \alpha}{d \tau} & = & \chi \frac{s(\tau)}{2} - \frac{\alpha}{2} \; \;. 
\end{eqnarray}

It is easy to express the expectation values of $\hat{x}, \hat{p}$ and others 
operators in terms of $\alpha$ and $\beta$ as follows;
\begin{eqnarray}
\langle \hat{x} \rangle & = & \frac{\alpha \beta + \alpha^* \beta^*}{\beta  + \beta^*} \; , \nonumber\\ 
\langle \hat{p} \rangle & = & -\frac{2i \beta \beta^* (\alpha-\alpha^*)}{\beta  + \beta^*} \; , \nonumber\\   
\langle \hat{x^2} \rangle & = & \frac{1}{\beta  + \beta^*} + {\langle \hat{x} \rangle}^2 \; , \nonumber\\  
\langle \hat{p}^2 \rangle & = & \frac{2 \beta \beta^*}{\beta  + \beta^*} -{\langle \hat{p} \rangle}^2  \; , \nonumber\\   
\langle \hat{L} \rangle & = & \frac{d \langle \hat{x^2} \rangle}{d \tau}  \; \; \;.
\end{eqnarray}

The wave function of Gaussian form (16) satisfies the minimum uncertainty 
condition
\begin{equation}
\Delta p \Delta x = \frac{1}{2}  \; \; \;.
\end{equation}
\noindent
with $\Delta x^2 = \langle \hat{x^2} \rangle - {\langle \hat{x} \rangle}^2$
and $\Delta p^2 = \langle \hat{p}^2 \rangle - {\langle \hat{p} \rangle}^2$.
For such a wave packet it is easy to see that
\begin{equation}
\beta = \beta^*  \; \; \;.
\end{equation}
\noindent
Since for a parabolic barrier the energy of classical particle is equal to the 
average energy of the quantum particle one can show that the energy spread for
the coupling-free parabolic barrier is
\begin{eqnarray}
\Delta \hat{H}_{q}^2 & = & \langle \hat{H}_{q}^2 \rangle-  \langle \hat{H}_{q} \rangle^2   \nonumber\\
& = & \omega_{0}^2 \left[ \langle \hat{p} \rangle^2 \Delta p^2 + \langle \hat{x} \rangle^2 \Delta x^2 \right] \; \; \;.
\end{eqnarray}

It is required that the minimum wave packet should have a minimum energy spread 
$\Delta \hat{H}_{q}^2$ also. Then we infer from Eq.(23) that the wave packet
must be prepared at the classical turning point $-x_{0}$, where 
$\langle \hat{p} \rangle = 0$ and $\langle \hat{x} \rangle^2$ has a minimum. 
Let us therefore assume that the center of the wave packet reaches the classical
turning point at $t=0$ with $\langle \hat{p} \rangle = p_{0} = 0$. We then have
[from $\hat{H}_{q} = \omega \frac{\hat{p^2}}{2}-\frac{\omega_{0} \hat{x}^2}{2}$, Eq.12]
\begin{equation}
\epsilon \omega_{0} = \frac{\omega_{0} x^2}{2}
\end{equation}
so that $x_{0} = \pm (2 \epsilon)^{\frac{1}{2}}$, where $\epsilon(=\frac{E_q}{\omega_{0}})$
is a dimensionless parameter (see Fig 1) denoting the energy of the particle
measured from the top of the barrier. We obtain
\begin{eqnarray}
\alpha(t=0) = \alpha_{0} = -x_{0} = -(2 \epsilon)^{\frac{1}{2}}  \nonumber\\
\beta(t=0) = \beta_{0} = \frac{1}{2} \; \; \;.
\end{eqnarray}
\noindent
The initial conditions (25) imply that the particle's wave packet is as compact
as possible when it arrives at the turning point. By choosing 
$\beta = \frac{1}{2}$, the wave packet satisfies the quantum-classical 
correspondence that the energy of the quantum particle is equal to that of the
classical particle for parabolic barrier.

The above two initial conditions for $\alpha$ and $\beta$ also set the initial
conditions for the others. Thus we further have from Eqs. (20),
\begin{eqnarray}
{\langle \hat{x} \rangle}_{t=0} & = & -(2 \epsilon)^{\frac{1}{2}} \; , \nonumber\\ 
{\langle \hat{p} \rangle}_{t=0} & = & 0 \; , \nonumber\\ 
{\langle \hat{x}^2 \rangle}_{t=0} & = & \frac{1}{2}(1+4 \epsilon) \; , \nonumber\\ 
{\langle \hat{p^2} \rangle}_{t=0} & = & \frac{1}{2} \; , \nonumber\\  
{\langle \hat{L} \rangle}_{t=0} & = & 0  \; \; \;. 
\end{eqnarray}
\noindent
Eqs. (26) suggest that the initial conditions required for the dynamics can be
manipulated by controlling a single parameter $\epsilon$, which denotes the
average energy of the wave packet with which the particle impinges on the
barrier at the left turning point.

Since the classical subsystem is coupled to the quantum degrees of freedom it 
is also necessary to specify the initial conditions for the classical variables
for its evolution. To this end we fix the parameter set $\omega = 6.32, \; A = 0.002,
\; B = 0.25 $ and $G =0.63 $ for the present study and choose the initial conditions 
for $s$ and $p_{s}$ for the regular and chaotic trajectories as $s= -4.02 , -4.52,
-5.02$ for three regular and $ s= -8.80,  -9.30, -11.31$ for three chaotic 
trajectories, $p_{s}$ being chosen to be zero always. We refer to Lin and
Ballentine [13] for a typical illustration of the phase space.

The initial conditions (25) and (26) along with those for the classical 
variables then allow us to follow the evolution of the expectation values and of 
the wave packet in terms of Eqs. (13-14) and Eq. 19. The relevant physical 
quantities of interest are determined in the next section.

Before closing this section a relevant point need to be discussed here.
We have considered the regular and chaotic trajectories of the classical
subsystem and referred to its phase space as if this subsystem were not 
coupled to the quantum subsystem. Since the quantum subsystem considered
here has a potential which is not bounded from below (a prototype model for 
barrier penetration), the phase space of the mixed system ( quantum plus 
classical) is unbounded and therefore the very notion of 
classical chaos is far from clear in such a situation. Based on this
consideration we have selected the chaotic and regular parts of the phase 
space corresponding to the classical subsystem only.
\vspace{0.5cm}
\begin{center}
{\bf
III. \hspace{0.2cm} The tunneling probability and the current}
\end{center}

We are now in a position to calculate the wave function and the corresponding
probability that the incident particle penetrates beyond the 
position $x=\zeta$. We write
\begin{equation}
T(x \geq \zeta, \tau) = \int_{\zeta}^\infty dx \left| \psi (x, \tau) \right|^2 \; \; \;.
\end{equation}

Normalization of the wave function(16) leads us to
\begin{equation}
\left|\psi(x, \tau) \right|^2 = \sqrt{\frac{a(\tau)}{\pi}} 
\; \exp[-a(\tau)(x-\langle x \rangle)^2] \; \; \; ,  
\end{equation}
\noindent
where $a(\tau) = \beta(\tau) + \beta^*(\tau)$. 

Since a portion of the wave packet has already tunneled through the barrier at
$\tau=0$, the tunneling contribution due to it should be subtracted from the
tunneling probability given by (27). Thus between $\tau=0$ and $\tau=\infty$
the wave packet tunnels with the probability
\begin{eqnarray}
p & = & T(x \geq x_{0}, \tau \rightarrow \infty) -  T(x \geq x_{0}, \tau = 0)  
\nonumber\\
& = & \sqrt{\frac{a(0)}{\pi}} \int_{0}^{2x_0} \exp [-a(0) y^2] dy 
- \sqrt{\frac{a(\infty)}{\pi}} \int_{0}^{x_{0} - \langle x \rangle_{\infty}} 
\exp[-a(\infty) y^2] dy  \; \; \;.
\end{eqnarray}

\noindent
Here $\langle x \rangle$ at $\tau=0$ is defined by  ${\langle x \rangle}_{0} = x_{0}$
and  $\langle x \rangle$ at $\tau= \infty$ by ${\langle x \rangle}_{\infty}$.

Another important quantity of interest is the tunneling current. The time 
evolution of the tunneling current
\begin{equation}
j(x, \tau) = \frac{1}{2i} [\psi^* \frac{\partial \psi}{\partial x}- (\frac{\partial \psi}{\partial x})^* \psi] \; \; ,
\end{equation}
\noindent
can be calculated by making use of the wavepacket(16) after normalization, together
with the solution for $\alpha(t)$ and $\beta(t)$ in terms of Eqs(18) and (19).
It is easy to see that the particle reaches the end of the tunnel 
at $x=x_{0}$ where it produces the following current:
\begin{equation}
j(x_{0}, \tau) = \frac{1}{i} [\beta \alpha - (\beta \alpha)^*-x_{0}(\beta-\beta^*)] {\left| \psi(x_{0}, \tau) \right|}^2  \; \; \;.
\end{equation}

We would now like to discuss here the following questions ;

(i) how the tunneling probability (29) depends on classical chaos due to the 
nonintegrability of the classical subsystem.

(ii) how long it takes a particle to tunnel through the barrier and how it depends
on classical chaos.

(iii) since the present model incorporates dissipation 
through classical friction, it is of interest to consider how the dissipative  
tunneling probability and the current depend on the nature of the classical
motion of the subsystem.

Let us first discuss the case of dissipation-free tunneling ($\Gamma=0$). Fig. 2
depicts the tunneling probability for two sets of regular and chaotic 
trajectories, when the incident wave packet carries the average energy 
$\epsilon = 1$ . It is evident that irrespective of the initial energy of the
classical subsystem, the tunneling probability is substantially higher for the 
chaotic trajectories than that for the regular one. The calculation is repeated 
for lower incident energy (i. e. higher $\epsilon$) of the wave packet 
($\epsilon = 3$) and the result is shown in Fig. 3. The difference is more 
marked in the region of higher values of coupling constant.

The tunneling current (31) is shown in Fig. 3 as a function of time for two
trajectories (one regular and the other chaotic). For a relative comparison we have
also plotted the current for the coupling-free case. We observe that a particle 
needs more time to tunnel through the barrier when the classical subsystem
is regular. 

We now turn to the case of dissipative tunneling ($\Gamma \ne 0$). In Fig. 4 we
display the effect of dissipation on the tunneling probability when the classical
subsystem is chaotic. 
A set of three chaotic trajectories corresponding to three 
different initial conditions ($p_s = 0.0;
s = -11.31, -9.30, -8.80$) which refer to the widely different energies of the
undriven oscillator were studied. Similarly the initial conditions for the set
of three regular trajectories corresponding to widely seperated turning
points of the undriven well ($p_s = 0.0; s = -5.02, -4.52, -4.02$) were
examined. A variation of $\Gamma$ was also carried out. For the 
sake of brevity we have plotted here only the  representative variation
for one chaotic ($s = -11.31, p_s = 0.0$) and one regular ($s = -5.02, p_s = 0.0$)
trajectory with and without dissipation.
It is  interesting to observe that dissipation increases
the tunneling probability quite significantly when the classical subsystem
is chaotic (solid curves) but for the regular classical subsystem (dotted curves)
tunneling probability is suppressed by dissipation.
Such a differential behavior of dissipative tunneling of wave packets 
for the chaotic and regular trajectories (which remains qualitatively the same 
for other sets) is also apparent in the peak height of the 
tunneling current (Fig. 5), although the tunneling time does not differ
too much in the two cases.

Summarizing the above numerical results we observe that (i) classical chaos of 
the subsystem increases the tunneling probability and decreases the tunneling 
time quite significantly. (ii) Dissipation enhances the tunneling when the
classical subsystem is chaotic in contrast to the case when the subsystem 
behaves regularly.

The problem of chaos-assisted tunneling has been addressed by various workers 
over many years. For example, while studying the model two-dimensional 
autonomous systems it has been observed that energy splitting can increase 
dramatically with chaos of the intervening chaotic layer in which tunneling 
takes place between distinct, but symmetry related regular phase space regimes 
separated by a chaotic layer [24]. Lin and Ballentine [17] carried out a numerical study
on a driven double-well oscillator to show that as the separating phase space 
layer grows more chaotic with increasing driving strength the tunneling rate is 
enhanced by orders of magnitude over the rate of the undriven system. Utermann
et. al. [15] also investigated the same system to point out the role of classical
chaotic diffusion as a mediator for barrier tunneling.

The enhancement of tunneling of wave packet by classical chaos in the present 
model of mixed quantum-classical description can be understood in the light of 
the of this classical chaotic diffusion. Instead of considering tunneling 
between two regions(tori) mediated by a chaotic transport between them we 
consider here a single tunneling process which starts at one of the turning 
points of the barrier. As the wave packet evolves through the barrier its mean 
motion by virtue of the coupling of the inverted potential with classical 
subsystem is affected by classical chaotic diffusive motion of the latter. 
While in the former cases tunneling is reversible, the process considered
here is irreversible in nature.

The role of dissipation is somewhat more intriguing since one is concerned here
with three interplaying aspects of evolution, e. g, tunneling, classical chaos and
dissipation. The mechanism of enhancement of tunneling probability by 
dissipation when the subsystem is chaotic, can be understood qualitatively in the
following way; It has been shown recently [19] that tunneling may get enhanced due to
virtual mixing of excited states by dissipative interaction in contrast to ground
state tunneling which is suppressed due to dissipation. 
The more relevant to the present work is an earlier observation by Ford, Lewis and
O'Connell [25] who had employed a quantum Langevin equation and found an 
increased tunneling rate for a particle tunneling through
a parabolic barrier in a black-body radiation(incoherent) field which behaves
as a standard harmonic bath. It is 
important to realize that in the present model the classical
chaotic subsystem although a few degree-of-freedom system mimics
the behaviour of a typical heat bath [22] which in course of energy exchange
with the quantum subsystem acquires partial quantum character and brings in 
decoherence in the dynamics.
The classical chaotic subsystem is thus reminiscent of 
the background of a black-body radiation field reservoir
and the enhancement of the resulting tunneling of the wave packet
through a parabolic barrier can be understood qualitatively in the
spirit of Ford, Lewis and O'Connell [25].

\vspace{0.5cm}

\begin{center}
{\bf
IV. \hspace{0.2cm} Conclusion}
\end{center}

In this study we have considered the tunneling of a Gaussian wave packet through
a potential barrier which  in turn is coupled to a nonintegrable classical 
system. The operators pertaining to this system with mixed quantum-classical
description close a partial Lie algebra with respect to the Hamiltonian 
operator of the system. By introducing a phenomenological classical friction
one realizes a mechanism of dissipation in the overall dynamical evolution
in this model without violating 
any quantum rule. Because of nonintegrability the classical subsystem admits of 
chaotic behavior. We have studied the interplay of classical chaos and
dissipation in the tunneling of wave packet through the barrier and shown
that chaos-assisted tunneling is further enhanced by dissipation while 
tunneling is suppressed by dissipation when the subsystem behaves regularly.
Dissipation thus plays a significant role in the evolution of a tunneling process 
in presence of classical chaos.

{\bf Acknowledgements:} BCB is indebted to the Council of Scientific and 
Industrial Research for partial financial support. DSR is thankful to the 
Department of Science and Technology for a research grant.

\newpage
\begin{center}
{\bf References}
\end{center}
\vspace{0.5cm}

\begin{enumerate}
\item P. M. Stevenson, Phys. Rev. {\bf D30} 1712(1984); {\bf D32} 1389(1985).
\item A. K. Pattanayak and W. C. Schieve, Phys. Rev. {\bf A46} 1821 (1992);
H. Schanz and B. Esser, Phys. Rev. {\bf A55} 3375 (1997).
\item P. W. Milonni, J. R. Ackerhalt and H. W. Galbraith, Phys. Rev. Letts. 
{\bf 50} 966 (1983); J. R. Ackerhalt and P. W. Milonni, J. Opt. Soc. Am. {\bf B1} 116 (1984);
P. W. Milonni, J. R. Ackerhalt and H. W. Galbraith, Phys. Rev. {\bf A28} 887 (1983);
P. I. Belobrov, G. P. Berman and G. M. Zaslavski, JETP
{\bf 49} 993 (1979).
\item A. Nath and D. S. Ray, Phys. Rev. {\bf A36} 431 (1987); Phys. Letts.
{\bf A117} 341 (1986); Phys. Letts. {\bf A116} 104 (1986); G. Gangopadhyay
and D. S. Ray, Phys. Rev. {\bf A40} 3750 (1989).
\item A. Bulgac, Phys. Rev. Letts. {\bf 67} 965 (1991).
\item R. Blumel and B. Esser, Phys. Rev. Letts. {\bf 72} 3658 (1994).
\item L. L. Bonilla and F. Guinea, Phys. Letts. {\bf B271} 196 (1991);
Phys. Rev. {\bf A45} 7718 (1992).
\item A. K. Pattanayak and W. C. Schieve, Phys. Rev. Letts {\bf 72} 2855 (1994).
\item A. M. Kowalaski, A. Plastino and A. N. Proto, Phys. Rev. {\bf E52} 165 (1995).
\item E. Kanai, Prog. Theo. Phys. {\bf 3} 440 (1948).
\item M. D. Kostin, J. Stat. Phys. {\bf 12} 145 (1975).
\item D. Greenberger, J. Math. Phys. {\bf 20} 762 (1979).
\item A. O. Caldeira and A. J. Leggett, Physica {\bf 121A} 587 (1983).
\item I. V. Krive and A. S. Rozhavskii, Theo. and Math. Phys. {\bf 89} 1069
(1991); I. V. Krive and S. M. Latinsky, Ann. Phys. {\bf 221} 204 (1993). 
\item W. H. Louisell, Quantum Statistical Properies of Radiation (Wiley, NY 1973).
\item D. S. Ray, Phys. Letts. {\bf A122} 479 (1987) ; J. Chem. Phys. {\bf 92}
1145 (1990) ; R. D. Levine and C. E. Wulfman Chem. Phys. Letts. {\bf 60}
372 (1974).
\item W. A. Lin and L. E. Ballentine, Phys. Rev. Letts. {\bf 65} 2927 (1990).
\item J. Plata and J. M. Gomez Llorente  J. Phys. {\bf A25} L303 (1992).
\item R. Utermann, T. Dittrich and P. H{\ae}nggi, Phys. Rev. {\bf E49} 273 (1994).
\item K. Fujikawa, S. Iso, M. Sasaki and H. Suzuki,
Phys. Rev. Letts. {\bf 68} 1093 (1992). See the references therein.
\item A. J. Leggett and A. O. Caldeira Phys. Rev. Letts. {\bf 46} 211 (1981).
\item S. Chaudhuri, G. Gangopadhyay and D. S. Ray, Phys. Rev. {\bf E52} 2262 (1995).
\item S. Chaudhuri, D. Majumdar and D. S. Ray, Phys. Rev. {\bf E53} 5816 (1996).
\item O. Bohigas, S. Tomsovic and D. Ullmo, Phys. Rep {\bf 223} 43 (1993).
\item G. W. Ford, J. T. Lewis and R. T. O'Connell, Phys. Letts. {\bf A158}  
367 (1991).
\end{enumerate}
\newpage
\begin{center}
{\bf Figure Captions}
\end{center}
\vspace{0.5cm}

Fig. 1. A gaussian wave 
packet is shown at the left classical turning point $-x_0$, of an inverted 
parabolic barrier.

Fig.2. The tunneling probability($p$) is plotted as a function of 
coupling strength ($\chi$), for the wave packet's average 
energy $E_{q}=\hbar \omega_{0} (\epsilon=3.0)$ for different
initial positions of classical subsystem. The intial positions for three classical chaotic
trajectories are ($a$) $s$=-11.31, ($b$) $s$=-9.30, ($c$) $s$=-8.80 
and for three regular trajectories and ($d$) $s$=-5.02, ($e$) $s$=-4.52 and 
($f$) $s$=-4.02; $p_s = 0$. (The quantities are dimensionless; scale arbitrary)

Fig.3. The tunneling current(T) is plotted as a function of time($\tau$) for 
the wave packet's  average
energy $\epsilon=1.0$, ($a$) for classical
chaotic subsystem, ($b$) for classical regular subsystem and ($c$) uncoupled 
system ($\chi$)=0.

Fig.4. The tunneling probability($p$) is plotted as a function coupling 
strength($\chi$) for the wave packet's average energy $\epsilon=3$
for different damping values ($a$) $\Gamma$=2.0 and ($b$) $\Gamma$=0.0 when
the classical subsystem is chaotic. 
Similarly curves (c) $\Gamma$=2.0 and 
(d) $\Gamma$= 0.0 are plotted for the regular classical subsystem.

Fig.5. The tunneling current(T) is plotted as a function time 
($\tau$) for the wave packet's average energy $\epsilon=3$
for different damping values ($a$) $\Gamma$=2.0 and ($b$) $\Gamma$=0.0 when
the classical subsystem is chaotic.
Similarly curves (c) $\Gamma$=2.0 and 
(d) $\Gamma$= 0.0 are plotted for the regular classical subsystem.

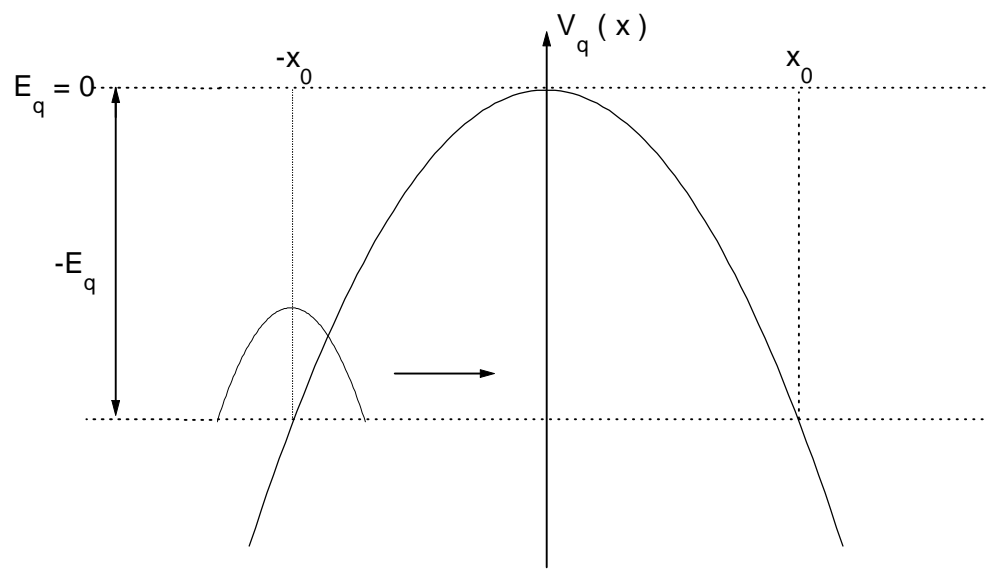
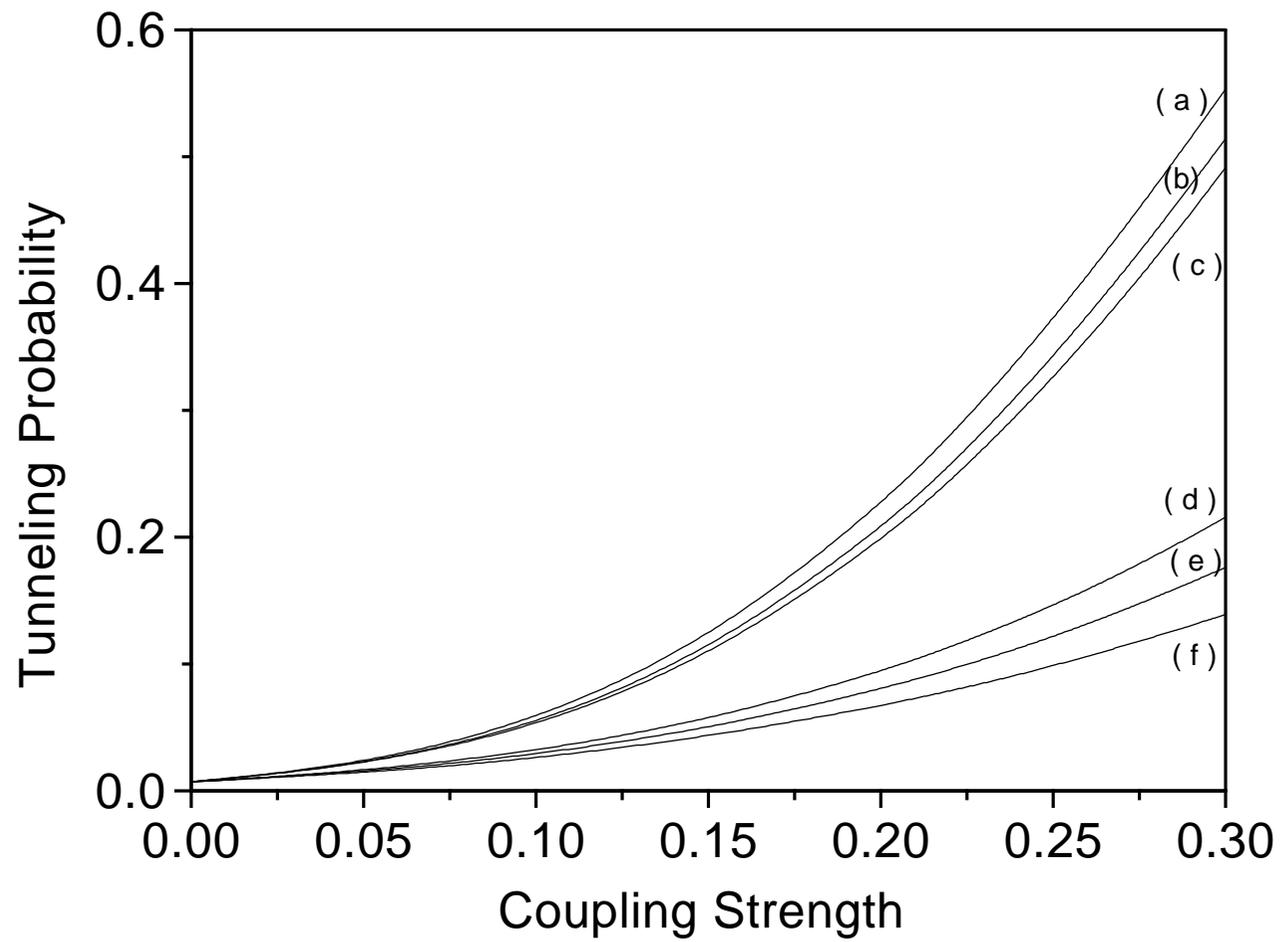
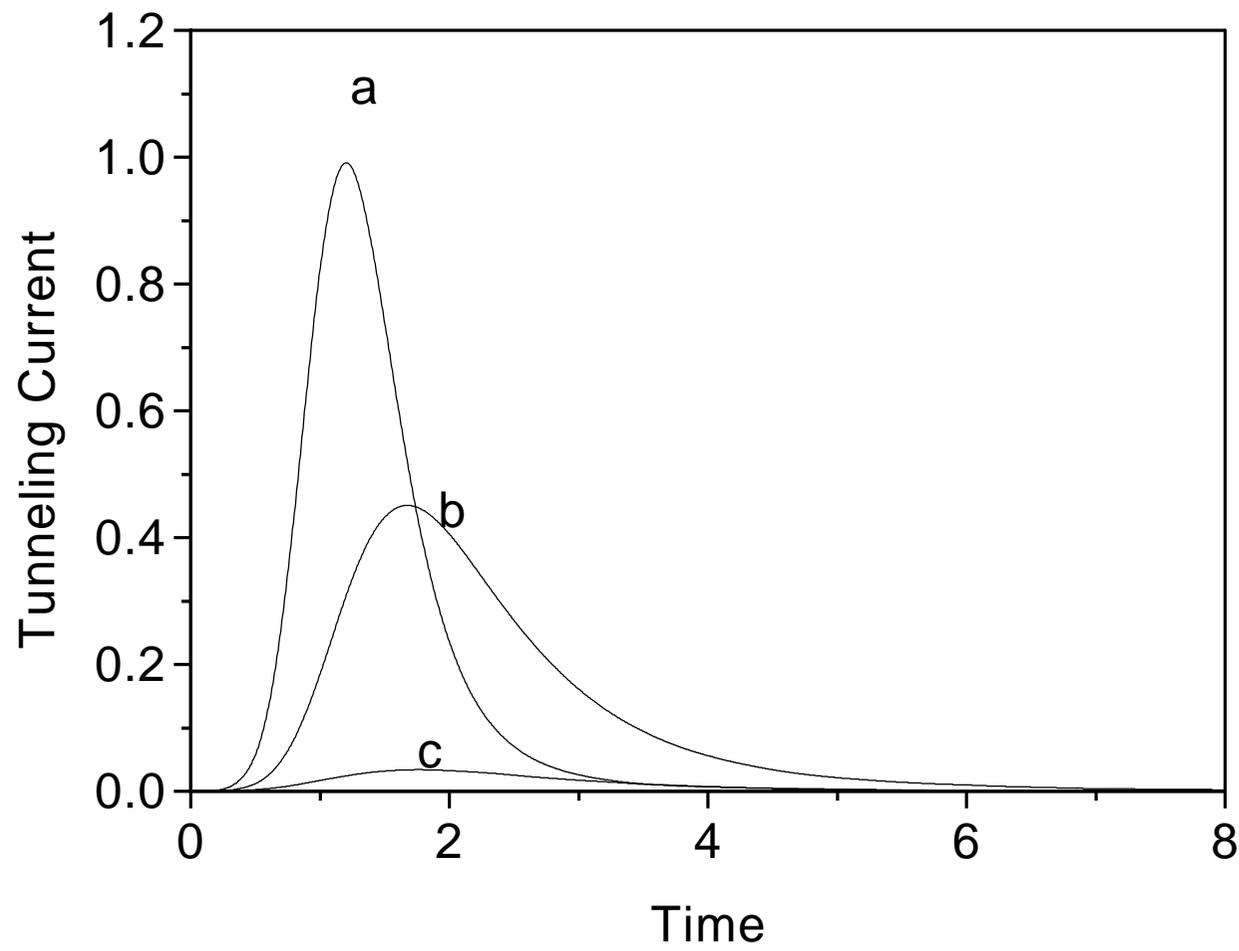
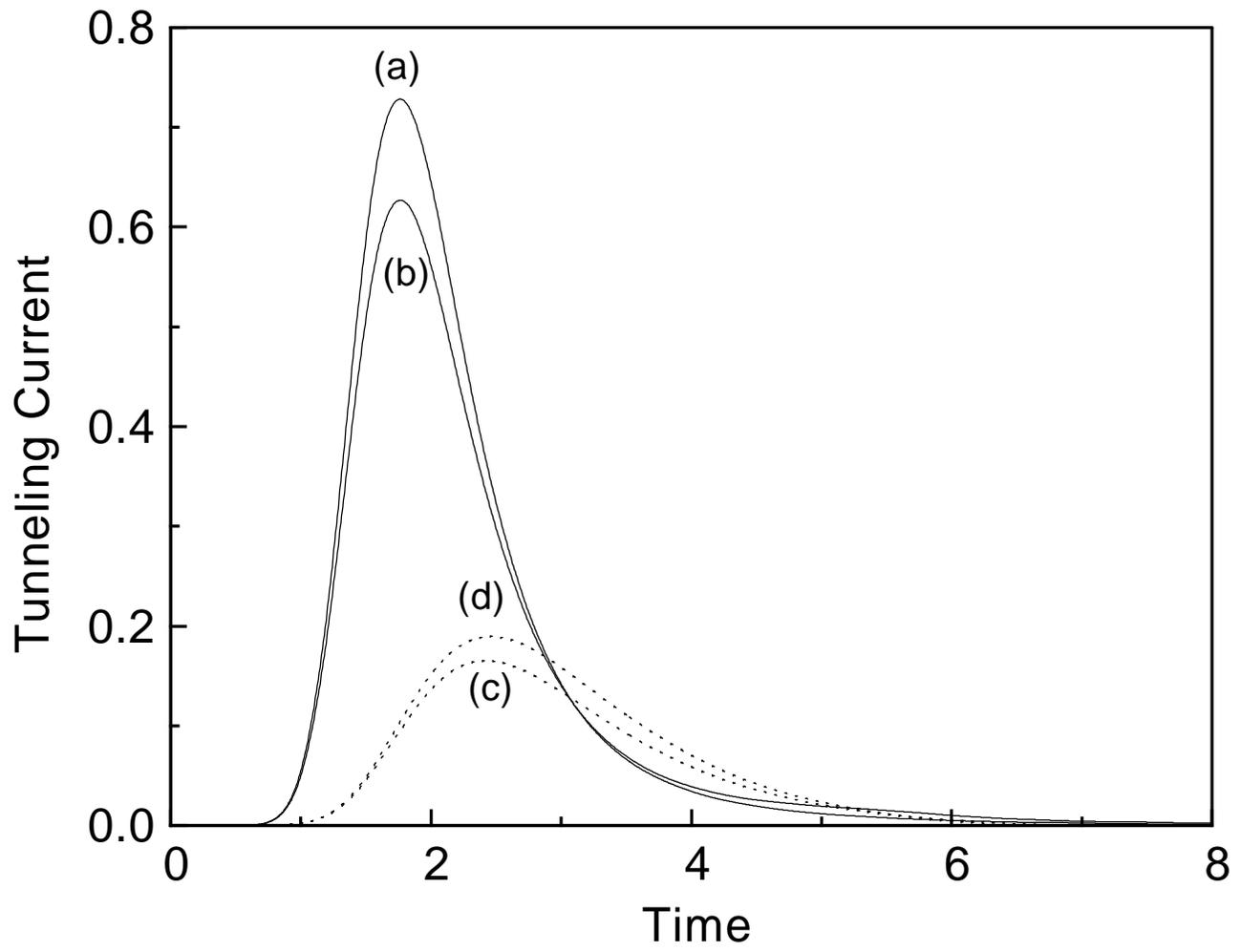
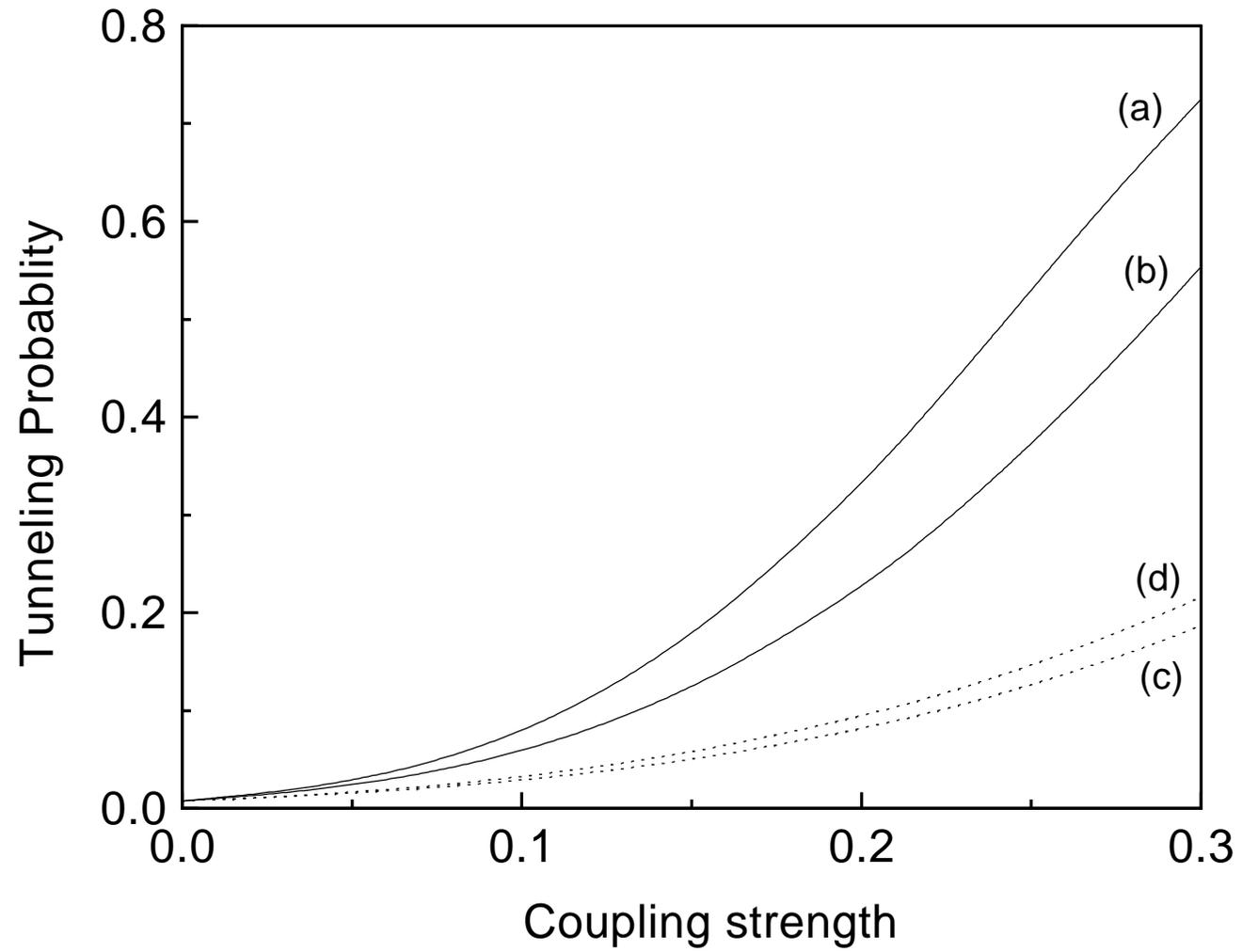
\end{document}